\documentclass[sor4-1,graphicx,unsortedaddress,preprint,longbibliography]{revtex4-1}
\usepackage{graphicx} 
\usepackage{amsmath,amssymb}
\usepackage{natbib}
\usepackage[english]{babel}
\usepackage[usenames,dvipsnames]{xcolor}

\draft % marks overfull lines with a black rule on the right

\begin{document}

% Use the \preprint command to place your local institutional report number 
% on the title page in preprint mode.
% Multiple \preprint commands are allowed.
%\preprint{}

\title{A Well Defined Glass State Obtained by Oscillatory Shear} %Title of paper

% repeat the \author .. \affiliation  etc. as needed
% \email, \thanks, \homepage, \altaffiliation all apply to the current author.
% Explanatory text should go in the []'s, 
% actual e-mail address or url should go in the {}'s for \email and \homepage.
% Please use the appropriate macro for the type of information

% \affiliation command applies to all authors since the last \affiliation command. 
% The \affiliation command should follow the other information.

\author{Lisbeth P{\'e}rez-Ocampo}
%\email[]{Your e-mail address}
%\homepage[]{Your web page}
%\thanks{}
%\altaffiliation{}
\affiliation{Divisi{\'o}n de Ciencias e Ingenier{\'i}as, Universidad de Guanajuato, Lomas del Bosque 103, 37150 Le{\'o}n, Mexico}

\author{Alessio Zaccone}
% Collaboration name, if desired (requires use of superscriptaddress option in \documentclass). 
% \noaffiliation is required (may also be used with the \author command).
%\collaboration{}
%\noaffiliation
\affiliation{Department of Chemical Engineering and Biotechnology, and Cavendish Laboratory, University of Cambridge, Cambridge CB3 0AS, UK}

\author{Marco Laurati}
\email[]{mlaurati@fisica.ugto.mx}
%\homepage[]{Your web page}
%\thanks{}
%\altaffiliation{}
%\noaffiliation
\affiliation{Divisi{\'o}n de Ciencias e Ingenier{\'i}as, Universidad de Guanajuato, Lomas del Bosque 103, 37150 Le{\'o}n, Mexico}
\date{\today}

\begin{abstract}
We investigate the process of shear melting and re-solidification of a colloidal glass, directly after loading (pre-yielding) and after a series of consecutive strain sweeps (post-yielding). The post-yielding glass shows a significant softening compared to the pre-yielding glass, together with the absence of history effects in successive shear melting protocols, indicating a reproducible process of fluidisation and re-solidification into a glass state unaffected by residual stresses. However, a significant hysteresis characterises strain sweeps with increasing or decreasing strain amplitude. The appearance of history and hysteresis effects coincides with the formation of a glass state, whereas it is not observed in the liquid. We can describe the onset of shear melting over a broad range of volume fractions and frequencies using a recently developed model which describes the yielding process in terms of loss of long-lived nearest neighbours.
\end{abstract}

% insert suggested PACS numbers in braces on next line

\maketitle %\maketitle must follow title, authors, abstract and \pacs

% Body of paper goes here. Use proper sectioning commands. 
% References should be done using the \cite, \ref, and \label commands
\section{Introduction}
%\label{}
Application of shear to a concentrated colloidal dispersion typically induces flow. The flow properties have important consequences for the processing of dispersions that are of intertest in applications, like paints, foods and drilling fluids, among others \cite{larson}. These properties depend on the characteristics of the dispersion, like the particle-particle interactions, shape, surface properties and size distribution, but also on the parameters of the applied shear field \cite{wagner}. For the latter, continuous or oscillatory shear might lead to different responses \cite{voigtmann2014,kobelev2005}, as well as application of strain or stress \cite{siebenburger,tatjana_scirep,tatjana_new,kobelev2005,rogers2011a}. Moreover, the control parameters, like the shear rate \cite{laurati2012}, oscillation frequency \cite{petekidis02}, applied stress \cite{tatjana_scirep,moraz14}, have important influence on the flow behaviour. Despite these studies, a comprehensive understanding of the influence of all these factors is yet to be obtained.
 
Often, model systems are used to investigate general properties of a class of materials. For colloids, a suitable model system is a dispersion of hard-sphere like particles \cite{poon_weeks}. At large particle volume fractions ($\phi\gtrsim 0.58$) these dispersions form a non-equilibrium amorphous solid-state, a glass, due to the dynamical arrest induced by crowding \cite{pusey,vanmegen}. Under application of a constant shear rate or stress, the solid melts and flows \cite{besseling07,Koumakis2012,koumakis2016,amann,schall_zaccone}. Glass melting has been explained in terms of the rearrangement of the cage structure surrounding a particle \cite{Koumakis2012,fuchs2015}, through non affine particle motions and plastic events \cite{Alessio_new}. Cage rearrangements have been also associated to negative stress correlations by Mode-Coupling Theory (MCT) \cite{zausch08,voigtmann2014}.

The application of oscillatory shear at constant frequency and with a sufficiently large strain or stress amplitude similarly leads to structural yielding and flow \cite{petekidis02,pham08,vanderwaart,koumakis2013}, also associated to cage breaking \cite{pham08,koumakis2013,koumakis2016} and to the onset of irreversible particle motions \cite{petekidis02,ganapathy,jeanneret,knowlton}. Shear melting due to cage breaking is also observed in softer glasses \cite{helgeson2007,rogers2011a,renou2010,koumakis2012b}, however accompanied to significant qualitative differences in the detailed yielding behavior \cite{koumakis2012b}. Most of the studies on hard-sphere glasses discuss the initial yielding and transition to flow when, starting from the state after loading (pre-yielding) or a reproducible state after pre-shearing, the strain or stress amplitudes are progressively increased (strain or stress sweep). Less is known about structure reformation when, starting from the flowing state achieved after yielding, the strain amplitude $\gamma$ is progressively decreased down to the linear response regime. Due to the presence of unrelaxed stresses in the flowing glass \cite{ballauff,cloitre}, history effects in glass reformation might lead to a different glass state (post-yielding), with distinct mechanical properties compared to the pre-yielding solid \cite{viasnoff}. Recent results on model glasses indicate the presence of these history effects and their connection to structural rearrangements \cite{schall_zaccone}. 

However, the process of shear-melting and re-solidification of the post-yielding glassy dispersion was not investigated in detail. In particular the question arises whether application of a series of consecutive shear-melting and re-solidifcation processes to the post-yielding glassy dispersion leads to ever different glass states, associated with the presence of residual stresses in the shear-molten solid. Alternatively, repeated processes of shear-melting and re-solidification might lead to a well defined solid state with reproducible mechanical properties. The ability of shear to induce structural arrangements facilitating flow, and the presence of memory effects, have been suggested by studies on non-Brownian dispersions subjected to several cycles of deformation \cite{keim,foffi,regev2013}. It should be noted that for more monodisperse samples than the ones studied here, the application of repeated cycles of oscillatory shear might lead to crystallization, as already observed for similar dispersions of hard spheres and colloid-polymer mixtures \cite{koumakis2008,smith07}. To investigate the effects of repeated cycles of shear-melting and re-solidification on the glass state, we study  the mechanical response of hard-sphere glasses and concentrated hard-sphere fluids to several consecutive forward (increasing $\gamma$) and backward (decreasing $\gamma$) dynamic strain sweeps.   We show that in the glass the post-yielding solid, while presenting significant softening compared to the pre-yielding solid, shows reproducible evolution of the moduli during the shear-melting and re-solidification processes. However, a reproducible and significant hysteresis is observed between the moduli measured in strain sweeps during shear melting (increasing strain amplitude) or re-solidification (decreasing strain amplitude). Moreover, by comparison with the fluid state, we demonstrate that the appearance of the initial softening of the modulus and hysteresis in the post-yielding glass are signatures of glass formation.\\
In order to demonstrate the qualitative differences between the pre-yielding and the post-yielding glass, and in particular the existence of a glass state with reproducible mechanical properties over several cycles of shear-melting and re-solidification, we present here measurements of the shear moduli in first harmonic approximation. More sophisticated analysis of large amplitude oscillatory shear data, taking into account higher harmonic contributions, exist \cite{mckinley,rogers2011b,ewoldt2013}: while they could provide additional information on the intra-cycle yielding, we expect that they would not change the qualitative results provided in the first harmonic approximation. We therefore postpone these analyses to later studies.\\
In addition, we rationalise the onset of shear melting for a broad range of colloid volume fractions and oscillations frequencies, in terms of the loss of long-lived neighbours, as recently proposed for the transition to flow of glasses under application of a constant shear rate \cite{Alessio_new}.

\section{Materials and Methods}

\subsection{Samples}

We investigated dispersions of polymethylmethacrylate (PMMA) particles sterically stabilised with  poly-hydroxystearic-acid (PHSA) of radius $R=150$\,nm and polydispersity of about 12\%, as determined by static and dynamic light scattering on a very dilute sample with $\phi < 10^{-3}$. 
The particles were suspended in a mixture of octadecene and bromonaphtalene to minimise solvent evaporation. Due to the relatively small size of the particles gravity effects were found to be negligible over the experimental measuring times. In this solvent mixture PMMA particles behave as nearly hard-spheres \cite{nick2011,royall}. 
Samples at different volume fractions were obtained by diluting a sediment obtained by centrifugation, for which we estimated a volume fraction of $\phi = 0.66$ according to simulation results \cite{Schaertl1994}. After dilution samples were homogenised in a rotating wheel for at least 1 day.

%\subsubsection{}
\subsection{Rheology}

Rheological measurements were performed using a DHR3 stress-controlled rheometer (TA Instruments) with a cone-plate geometry having a diameter of 50~mm and a cone angle of 0.5$^{\circ}$. A solvent trap was used to minimise solvent evaporation. In order to avoid the occurrence of wall slip, the geometries were spin-coated with a $\phi = 0.35$ dispersion of larger ($R_B = 720$ nm) PMMA spheres. The deposited layer of particles was then synthered at T = 110 $^{\circ}$C for 1 hour \cite{ballesta2012}.
For each sample, several consecutive Dynamic Strain Sweeps (DSS) were performed for different frequencies, alternating increasing (forward) and decreasing (backward) strain amplitude $\gamma$. The forward DSS spanned a range of amplitudes $10^{-3}\le \gamma\le 10$. Each backward DSS was started from the maximum $\gamma$ achieved in the previous forward test  ($\gamma = 10$) and $\gamma$ was reduced progressively down to $\gamma = 10^{-3}$. Before a series of repeated DSS at one frequency, a rejuvenation procedure was performed, such that each series started from a reproducible state of the sample. On the other hand, no rejuvenation was performed in between the consecutive DSS tests of a series and all repeated measurements were performed on the same sample loading. 
The rejuvenation procedure consisted of a step rate test, i.e. application of a step of constant shear rate $\dot{\gamma} = 0.1 $~s$^{-1}$ deformation until the steady state of flow was reached. After that we performed a Dynamic Time Sweep (DTS) with $\gamma = 0.01-0.1$\,\% (depending on the sample) in the linear viscoelastic  regime, extended until the elastic, $G'$, and viscous, $G''$, moduli reached constant steady-state values.

\subsection{Phenomenological model}

The model we use in this manuscript describes the deviation from the linear response regime and the onset of shear melting in dense fluids and glasses. It has been described in detail in recent work \cite{schall_zaccone,Alessio_new}. We recall here the main features and assumptions of the model. We remark also that the model in its present state cannot provide insights into the hysteresis and oscillatory steady state responses discussed later in the manuscript.
 
\subsubsection{Onset of Yielding}

In the model of shear-melting the decrease of G$^\prime$ with increasing strain amplitude $\gamma$ is the result of shear-induced loss of long-lived nearest neighbours. The long-lived neighbours are only a fraction of the total number of nearest neighbours that can be measured in a snapshot. A significant number of the nearest neighbours is indeed continuously changing due to fast and large oscillations induced by shear, and therefore cannot provide a significant contribution to stress transmission.

When shear is applied to a glassy dispersion of colloidal particles, the long-lived nearest neighbors constituting the cage of any tagged particle tend to leave the cage in the extensional sectors, whereas almost no new long-lived neighbors move in along the compression sectors as a result of excluded volume~\cite{Zausch09,Chikkadi12,Zaccone_prb}. This results in a negative balance of long-lived, mechanically-active nearest-neighbours which leads to a weakening of the stress-bearing structure of the glass with increasing strain amplitude~\cite{Zaccone11}. This effect has been confirmed experimentally at the microscopic level in recent work for a strain ramp in start-up shear~\cite{Alessio_new}. In ~\cite{Alessio_new} the number of long-lived nearest neighbors was experimentally determined as a function of accumulated strain and used to calculate the stress-strain response of the system.  The loss of long-lived nearest neighbors leads to increasing non-affine displacements until, at yielding, the effective number of nearest neighbors is barely enough to sustain the nonaffine displacements required to keep mechanical equilibrium~\cite{Zaccone11}. This picture is also consistent with the "percolation" of plastic events in the sample.
%while no energy is left to react to the deformation.
%This point defines the transition from solid to liquid at which the material starts to flow.

As previously shown \cite{schall_zaccone,Alessio_new}, the affine part of the shear modulus can be written as $G^{\prime}_A=\frac{1}{5\pi}\frac{\kappa\phi}{R}n(\gamma)$ in the linear regime. Here, $n(\gamma)$ is the number of long-lived nearest neighbors. The elastic spring constant $\kappa$ is defined as $\kappa=[d^{2}V_{\mathrm{eff}}/dr^{2}]_{r=\sigma}$, where $V_{\mathrm{eff}}/k_\mathrm{B}T=-\ln g(r)$ is the potential of man force between two bonded neighbors. The number of bonded neighbors can be obtained from the integral of the first peak of $g(r)$, which yields $n_{0} \approx 12$ for the static hard-sphere glass, as verified in~\cite{Alessio_new}. As mentioned, under applied shear particles become crowded in the compression sector of the shear plane, whereas particles become dilute in the extension sector, where long-lived neighbours are lost.

Recent results show that the number of nearest neighbors decreases exponentially with strain amplitude in large amplitude oscillatory shear experiments \cite{schall_zaccone}, $n(\gamma) = (n_{0}-n_c)\exp(-\alpha\gamma)+n_c$, or super-exponentially under application of a step to a constant shear rate \cite{Alessio_new}. To be consistent with previous work on large amplitude oscillatory shear, we choose here to use the exponential dependence on $\gamma$. The numerical factor $\alpha$ in the exponential decay is determined in the fitting to experimental data. Since this parameter represents the extent of the shear-induced microscopic connectivity loss, its fitted values, as discussed below, may vary depending on the shear protocol, the glass volume fraction, and the frequency. The parameter $n_{c}$ is the critical number $n_{c}=6$ of long-lived neighbors for central-force interactions~\cite{Zaccone11}. Once the system has become fluidlike, a finite $n_{c}=6$ is expected due to the steady-state hydrodynamic flow and its local structure \cite{brady_morris_1997}. This flow pushes the six neighbors in compression direction towards the particle and thus they remain for a long time.

As a result of the reduced connectivity, there are increasing non-affine contributions to the shear modulus, as shown recently also in numerical simulations~\cite{Priezjev}. According to previous work \cite{schall_zaccone,Alessio_new}, the non-affine contribution to the shear modulus is defined as $G'_{NA}$ in $G'=G'_{A}-G'_{NA}$ and can be written as 
$G'_{NA}= \frac{1}{5\pi}\frac{\kappa\phi}{R}~n_{c}$. This value results from the fact that while the affine part is proportional to the total number of mechanical constraints $n$, the non-affine part is instead related to the relaxation of local forces that arise due to the local lack of inversion symmetry in a disordered solid. Hence, given its nature of relaxation process, the non-affine contribution is proportional to the total number of degrees of freedom, $3N$. Upon factoring out common pre-factors, this leaves the well known scaling $G\sim (n-6)$, as derived with full details in Ref.~\cite{Zaccone11}.
Combining the affine and non-affine contributions we can therefore write the storage modulus as:

\begin{equation}
 \mathrm{G}^{\prime} = \mathrm{G}^{\prime}_{A}-\mathrm{G}^{\prime}_{NA}= K [(n_{0}- n_{c})\exp(-\alpha\gamma)]
   \label{Gprime}
\end{equation}

This expression, as already mentioned, can describe the initial deviation from linear behavior of $\mathrm{G}^{\prime}(\gamma)$, for $\gamma < \gamma_c = 1/\alpha$. The prefactor $K=\frac{1}{5\pi}\frac{\kappa\phi}{R}$.

\subsubsection{\label{theory_large}Large strain amplitude regime}

The model detailed in the previous section is able to  describe the initial regime of yielding of the solid, i.e. the initial decay from the linear response regime, which is also poorly affected by anharmonic contributions to the shear moduli \cite{mckinley,rogers2011b,ewoldt2013}. The transient regime after the initial yielding, in which the system is not completely fluidized and $\gamma\approx\gamma_c$ is not described by the model. At large strains $\gamma\gg\gamma_c$ previous work has shown that in the flow regime G$^{\prime}$ and G$^{\prime\prime}$ follow a power-law dependence on $\gamma$ which is characteristic of shear thinning \cite{brader,nick2011,miyazaki}. We therefore analyse, in addition to the initial yielding regime, the regime of large strain amplitudes by fitting the data with an empirical power-law dependence on strain amplitude $\gamma$, $G^{\prime} =  B\gamma^{\nu}$.

%\subsubsection{Storage modulus}
%As commented above, our moBy bringing the various contributions together, we finally arrive at the full expression for the storage modulus used in the fitting:
%\begin{equation}
% \mathrm{G}^{\prime} = 
%  \begin{cases} 
%   \mathrm{G}^{\prime}_{A}-\mathrm{G}^{\prime}_{NA}= K [n_{0}\exp(-\alpha\gamma) - n_{c}] &  \gamma\le \gamma_t \\
%   B\gamma^\nu       & \gamma > \gamma_t
%   \label{Gprime}
%  \end{cases}
%\end{equation}
%The characteristic strain amplitude $\gamma_t$ corresponds to the transition from the initial yielding regime to the flow regime, and is quantified by the value of $\gamma$ at which $n_{0}\exp(-\alpha\gamma) - n_{c}=0$. For $\gamma> \gamma_t$, $n_{0}\exp(-\alpha\gamma) - n_{c}$ would become negative, which is an unphysical condition.  Therefore the reduction of nearest neighbours corresponds to $n_0-n(\gamma_c) = 6$, in agreement with previous results \cite{Alessio_new}. 
%
% The fitting parameters used in the comparison with the data are $K$, $\alpha$, $B$ and $\nu$. While $K$ and $\alpha$ are constrained within reasonable ranges, $B$ and $\nu$ are freely adjustable parameter. As already mentioned, we assume $n_0=12$ in agreement with previous results.\cite{schall_zaccone,Alessio_new} The transition between the regimes with $\gamma > \gamma_t$  and $\gamma < \gamma_t$ is smooth for all fits. 

\section{Results and Discussion}

\subsection{Initial Shear-melting}
\subsubsection{Experiments}

\begin{figure}[h]
\centering
  \includegraphics[scale=0.4]{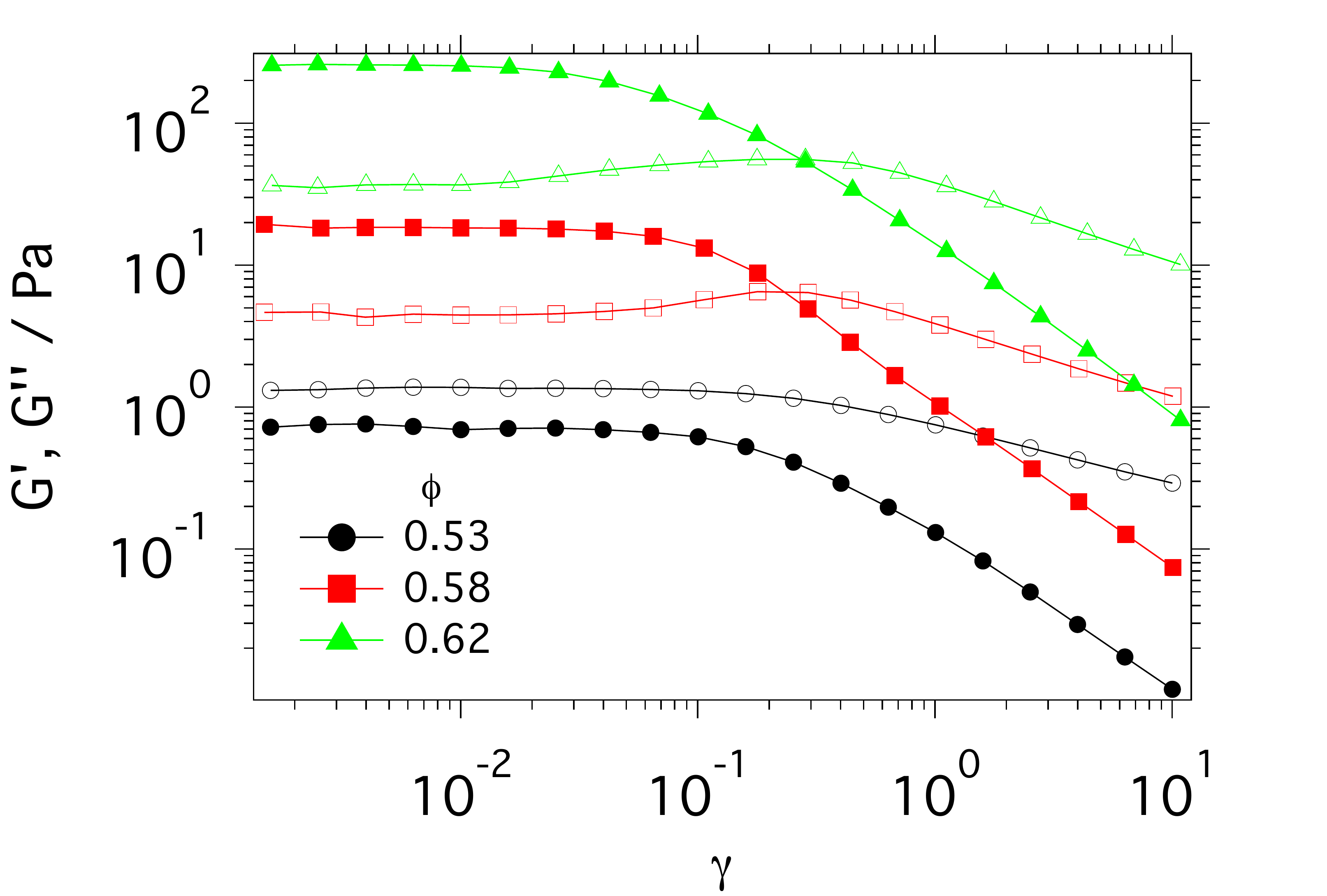}
  \caption{Storage (G$^{\prime}$, closed symbols) and Loss (G$^{\prime\prime}$, open symbols) moduli as a function of strain $\gamma$, obtained by Dynamic Strain Sweeps at frequency $\omega = 1$~rad/s for samples with different volume fractions $\phi$, as indicated.}
  \label{fig1}
\end{figure}

\begin{figure}[h]
\centering
  \includegraphics[scale=0.4]{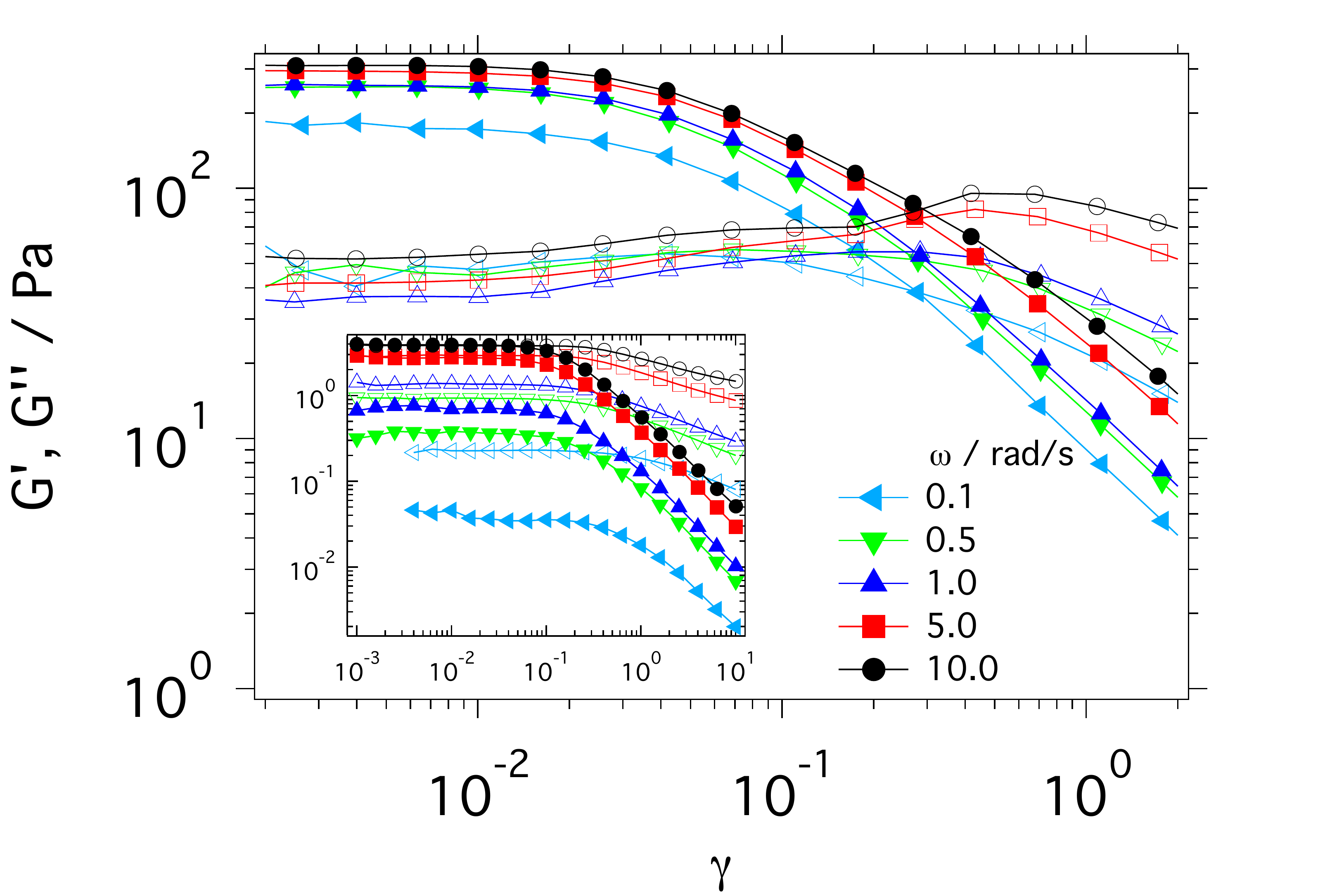}
  \caption{Storage (G$^{\prime}$, closed symbols) and Loss (G$^{\prime\prime}$, open symbols) moduli as a function of strain $\gamma$, for $\phi = 0.62$, obtained by Dynamic Strain Sweeps at different frequencies, as indicated. Inset: Comparable data for $\phi = 0.53$.}
  \label{fig_dss_freq}
\end{figure}

We investigated samples in a volume fraction range 0.53 $\lesssim\phi\lesssim$ 0.62, going from a fluid state at the smallest $\phi$ to a solid amorphous state at the highest $\phi$. The first DSS tests of a series (DSS 1) at $\omega = 1$~rad/s clearly show the transition between a fluid-like response and a solid-like response (Fig.\ref{fig1}) with increasing $\phi$: At $\phi = 0.53$ the loss modulus G$^{\prime\prime}$ is larger than the storage modulus G$^{\prime}$ for all $\gamma$, as expected for a fluid, and a power law dependence of the moduli close to that of a Newtonian fluid, i.e. G$^{\prime}\sim \gamma^2$ and G$^{\prime\prime}\sim\gamma$ is observed at large $\gamma$. At $\phi = 0.58$, where the glass transition of slightly polydisperse hard-spheres is expected \cite{vanmegen}, a solid-like response is observed in the linear viscoelastic regime, with G$^{\prime} > G^{\prime\prime}$, before the system yields at $\gamma\approx 0.2$, as indicated by the crossing of the storage and loss moduli and the maximum of G$^{\prime\prime}$ \cite{pham08}. This characteristic yield strain is associated to the maximum elastic deformation of a cage \cite{pham08}. The initial rise of G$^{\prime\prime}$ from the linear regime value up to a maximum is due to the rise of the viscous dissipative part of the response, while the purely elastic part ($G'$) decreases due to nonaffinity. The successive decrease after the peak can be interpreted as the consequence of the transition from the dominance of viscous/dissipative response into a new regime where dissipation is comparatively reduced due to shear-induced ordering manifested in a shear thinning regime, recently proposed for continuous shearing \cite{Alessio_new}. The system starts then to flow at larger $\gamma$. Fig.\ref{fig_dss_freq} shows the comparison between DSS tests at different frequencies of the glass sample with $\phi = 0.62$. The moduli increase in magnitude with increasing frequency, as expected since we are observing the system at increasingly shorter times where the response is increasingly solid. Moreover, at $\omega =5$ and 10 rad/s we observe two peaks in G$^{\prime\prime}$: One small peak at strain amplitudes $\gamma\approx 0.1$, which corresponds to the cage rupturing process, and a more pronounced peak at $\gamma\approx 0.4$, which has been associated with the occurrence of shear-induced collisions at high frequencies \cite{koumakis2013}. Finally, we can notice that the departure from the linear response regime appears to be approximately independent of frequency. The inset of Fig.\ref{fig_dss_freq} shows comparable results of the liquid sample with $\phi = 0.53$. Here we note also that the moduli approach each other at large frequencies in the linear regime, indicating that the liquid responds almost as a solid at short times. Furthermore we observe that the departure from the linear response regime here shifts to smaller $\gamma$ with increasing $\omega$, which can be understood based on the competition between Brownian relaxation and the timescale of shear, as will be discussed in more detail later. These responses are in good agreement with previous results on hard-sphere repulsive glasses \cite{pham08,schall_zaccone,miyazaki,brader}.

\subsubsection{Phenomenological model}

\begin{figure}[h]
 \centering
 \includegraphics[scale = 0.5]{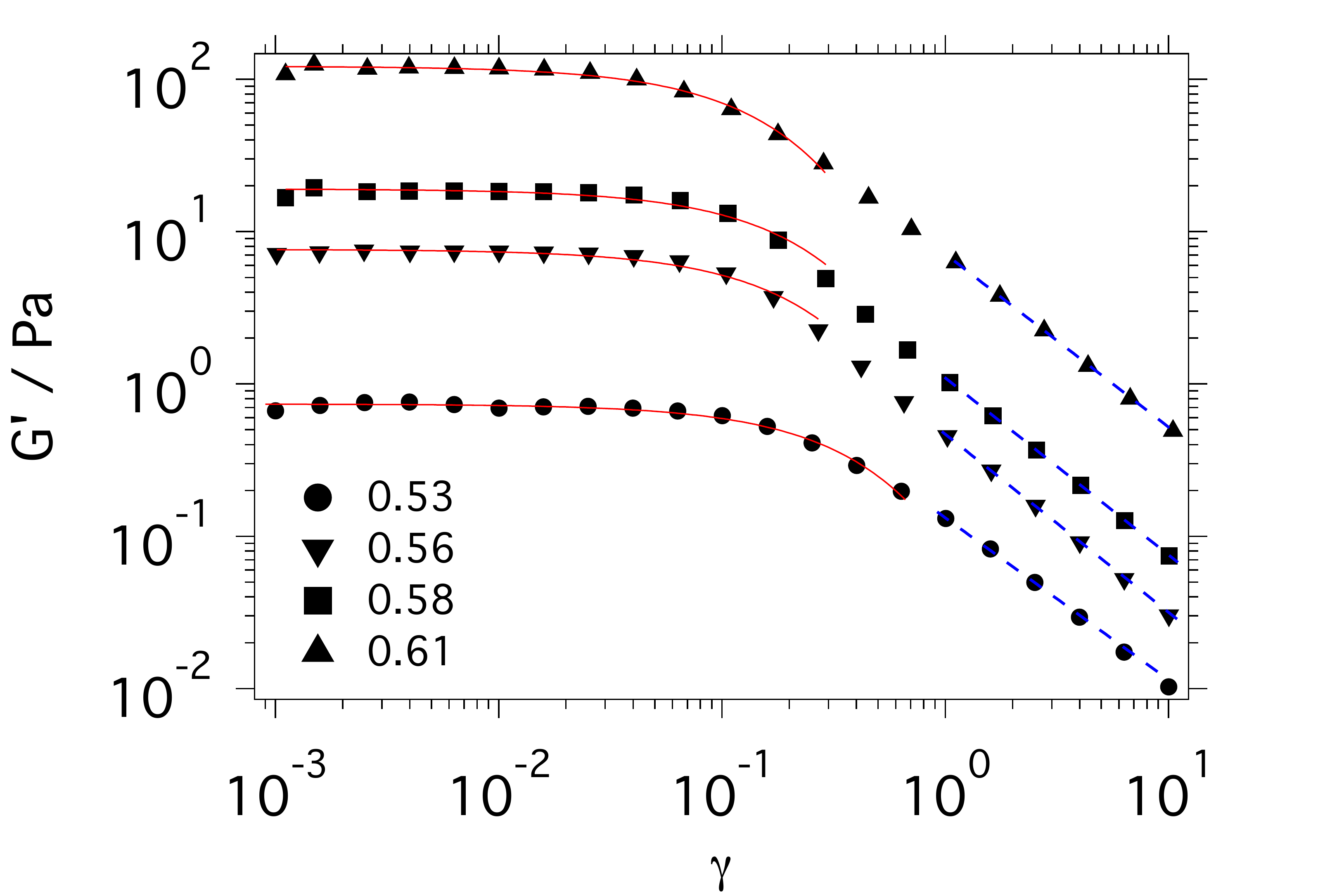}
 \caption{Exemplary model fits (red lines) of the storage (G$^{\prime}$, closed symbols) moduli of samples with different $\phi$, as indicated, as a function of strain amplitude $\gamma$, at $\omega = 1$~rad/s, obtained in the initial DSS measurements (DSS 1). Dashed blue lines represent power-law fits $G^{\prime} \sim \gamma^{\nu}$ of the final relaxation at large $\gamma$.}
 \label{fig6}
\end{figure}

We present in Fig.\ref{fig6} exemplary fits of the intial decay from the linear response regime of the storage moduli. The regime of small strain amplitudes, up to the initial deviation from the linear viscoelastic regime, which is quantified as the strain amplitude at which the storage modulus has decayed to 30\% of the value in the linear response regime, is fitted using the model of Eq. \ref{Gprime} (solid lines), with $K$ and $\alpha$ as free parameters. Note that the fit results would not change significantly if the last point of the fitted range is moved to the previous or next value of $\gamma$. Moreover the changes would only affect the absolute values of the parameters and not the trends reported in Fig. \ref{fig6}. In addition Fig.\ref{fig6} shows power-law fits of the final relaxation regime at large strains (dashed lines), as discussed in Section \ref{theory_large}.\\

{\bf Elastic Constant}\\

\begin{figure}[h]
 \centering
 \includegraphics[scale = 0.5]{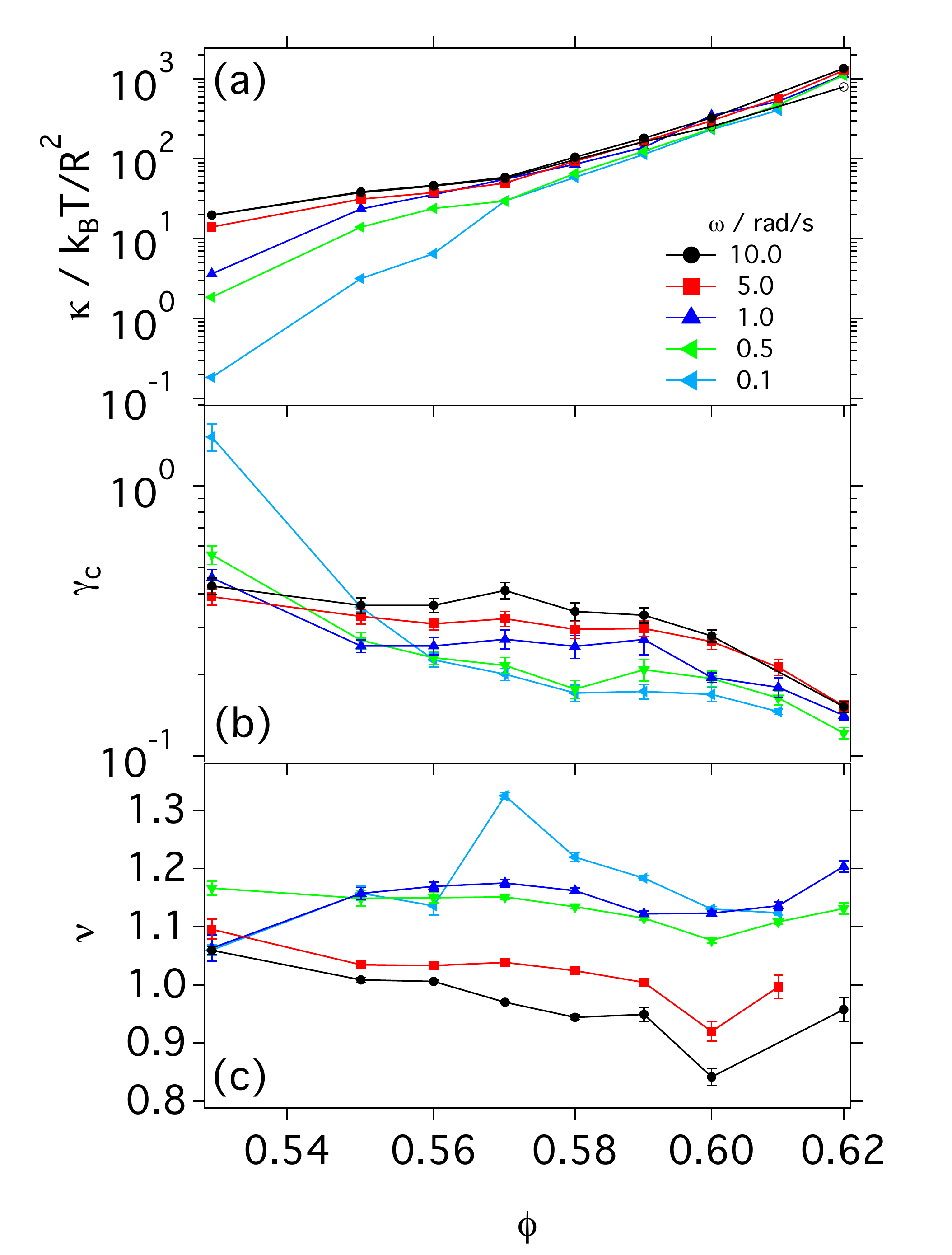}
 \caption{(a) Elastic spring constant $\kappa$, extracted from the prefactor $K$ obtained from fitting Eq. \ref{Gprime} to the experimental data, as a function of $\phi$, for different frequencies, as indicated. (b) $\gamma_c=1/\alpha$, obtained from fitting Eq.  \ref{Gprime} to the initial experimental DSS measurements, as a function of $\phi$, for different frequencies $\omega$ (same as in a). (c) Exponent $\nu$ obtained from power-law fits of the DSS data at large $\gamma$ for different frequencies (same as in a). Uncertainties, when larger than symbol sizes, are indicated by error bars.}
 \label{fig6b}
\end{figure}
%, as well as DSS 2 (not shown). 

We report in Fig.\ref{fig6b}a the elastic constant $\kappa = 5\pi KR/\phi$, obtained from the fitted values of $K$, for all values of the oscillation frequency and volume fraction $\phi$. We can observe that $\kappa$ increases with increasing $\phi$, in agreement with the stiffening of the dispersions when approaching and entering the glass state. At fixed frequency, the values of $\kappa$ are in good agreement with recent results on a similar system under continuous shearing \cite{Alessio_new}. It is interesting to note the presence of two clear regimes in the $\phi$ dependence of $\kappa$, separated by the value $\phi= 0.57$. The regime for $\phi < 0.57$ shows a marked frequency dependence and a strong $\phi$ dependence at small frequency, which becomes increasingly less pronounced and very weak at the highest frequency. On the other hand for $\phi > 0.57$ the $\phi$ dependence is moderate and similar at all frequencies. This different behavior in the two volume fraction regimes might depend on the fact that for the liquid at $\phi < 0.57$ the timescale of Brownian motion still competes with the timescale imposed by shear, while for the glass the timescale of Brownian motion becomes extremely large and therefore the timescale of shear dominates. Indeed we can estimate the dressed oscillatory Peclet number $Pe_{\omega} = \omega\tau_\mathrm{R}(\phi)$, with $\tau_\mathrm{R}$ the structural relaxation time, which estimates the relative contribution of Brownian and shear-induced relaxation, i.e.  $Pe_{\omega}\lesssim 1$ indicates that the structural relaxation is faster than the characteristic timescale of shear and viceversa. We use data on the volume fraction dependence of the long-time diffusion coefficient $D_\mathrm{L}(\phi)=f(\phi)D_0$ from the work of Van Megen and coworkers \cite{vanmegen98} to estimate $f(\phi$) and thus the structural relaxation time $\tau_\mathrm{R}(\phi)=R^2/D_\mathrm{L}(\phi)=6\pi\eta R^3/f(\phi)k_BT$. We obtain that $Pe_{\omega} \approx 0.9$ for the liquid at $\phi= 0.53$ and $\omega = 1$ rad/s, but becomes $>$ 1 for bigger values of $\omega$. Increasing $\phi$,  $Pe_{\omega}$ becomes $>$ 1 also at the smallest frequency and in the glass $Pe_{\omega} \gg 1$ for all frequencies.\\

{\bf Onset of Yielding}\\

In Fig.\ref{fig6b}b we show the dependences on $\phi$ and frequency of $\gamma_c = 1/\alpha$, with $\alpha$ the exponent of the exponential decay of the number of long-lived neighbours in Eq. \ref{Gprime}: $\gamma_c$ can be interpreted as a measure of the onset of \textit{anelasticity}, or onset of the departure from the linear response regime.
At small frequencies, up to 1 rad/s, $\gamma_\mathrm{c}$ is found to decrease with increasing $\phi$ in the fluid state. This indicates that the departure from the linear response regime occurs at increasingly smaller strain amplitudes when approaching the glass. This can be associated to the fact that in the fluid, Brownian motion gives flexibility to the cage, which can adapt to the deformation before breaking. With increasing $\phi$ the dynamics become slower: the cage becomes more rigid and can support less deformation before rearranging irreversibly. Subsequently, in the glass, $\gamma_\mathrm{c}$ remains approximately constant up to about $\phi = 0.60$, and then eventually decreases again. This approximately constant value can be associated with the size of the cage of nearest neighbours. The second decrease of $\gamma_\mathrm{c}$ can be associated with the approach to random close packing, i.e. the reduction of free volume in the system with the consequence that less and less space is available for rearrangements to sustain the deformation, and therefore the structure breaks at increasingly smaller $\gamma$.\\
With increasing frequency, the flat region extends to increasingly smaller $\phi$. When increasing $\omega$, the relaxation time of the fluid becomes increasingly large compared to the characteristic time of shear, as demonstrated earlier through the estimated values of the dressed oscillatory Peclet number $Pe^{\omega}$. Therefore Brownian motion, which helps in adapting the cage structure to the deformation before disruption, becomes increasingly irrelevant and the cage also in the fluid appears frozen on the timescale of the oscillatory deformation. Hence in the limit of high frequencies the fluid resembles the glass.\\ 

%\begin{figure}[h]
% \centering
% \includegraphics[scale = 0.38]{Fig5_new_powexp.pdf}
% \caption{Exponent $\beta$ obtained from power-law fits of the DSS data at large $\gamma$ for different frequencies, as indicated.}
% \label{fig7}
%\end{figure}
{\bf Final relaxation}\\

We finally report the values of the exponent $\nu$ obtained by fitting the final relaxation of G$^{\prime}$ with a power-law dependence. As shown in Fig.\ref{fig6b}c, $\nu$ presents values which are considerably smaller than 2, the value expected for the generalized Maxwell model \cite{brader}. Values considerably smaller than 2 for glasses have been observed before in experiments, simulations and theory \cite{brader,nick2011,miyazaki}. At the smallest frequency $\omega = 0.1$~rad/s, $\nu$ presents a maximum around $\phi = 0.57$. The origin of this maximum is not clear and the trend might be affected by the more pronounced noise of the experimental data at $\omega = 0.1$~rad/s and the smallest $\phi$ values. For intermediate frequencies, $\omega = 0.5$ and 1~rad/s, the exponent is approximately constant. For the highest frequencies, $\nu$ eventually slightly decreases with increasing $\phi$. The origin of these trends is not clear at present.

\subsection{Re-solidification and successive melting of the post-yielding dispersion}

\begin{figure}[t]
 \centering
 \includegraphics[scale = 0.5]{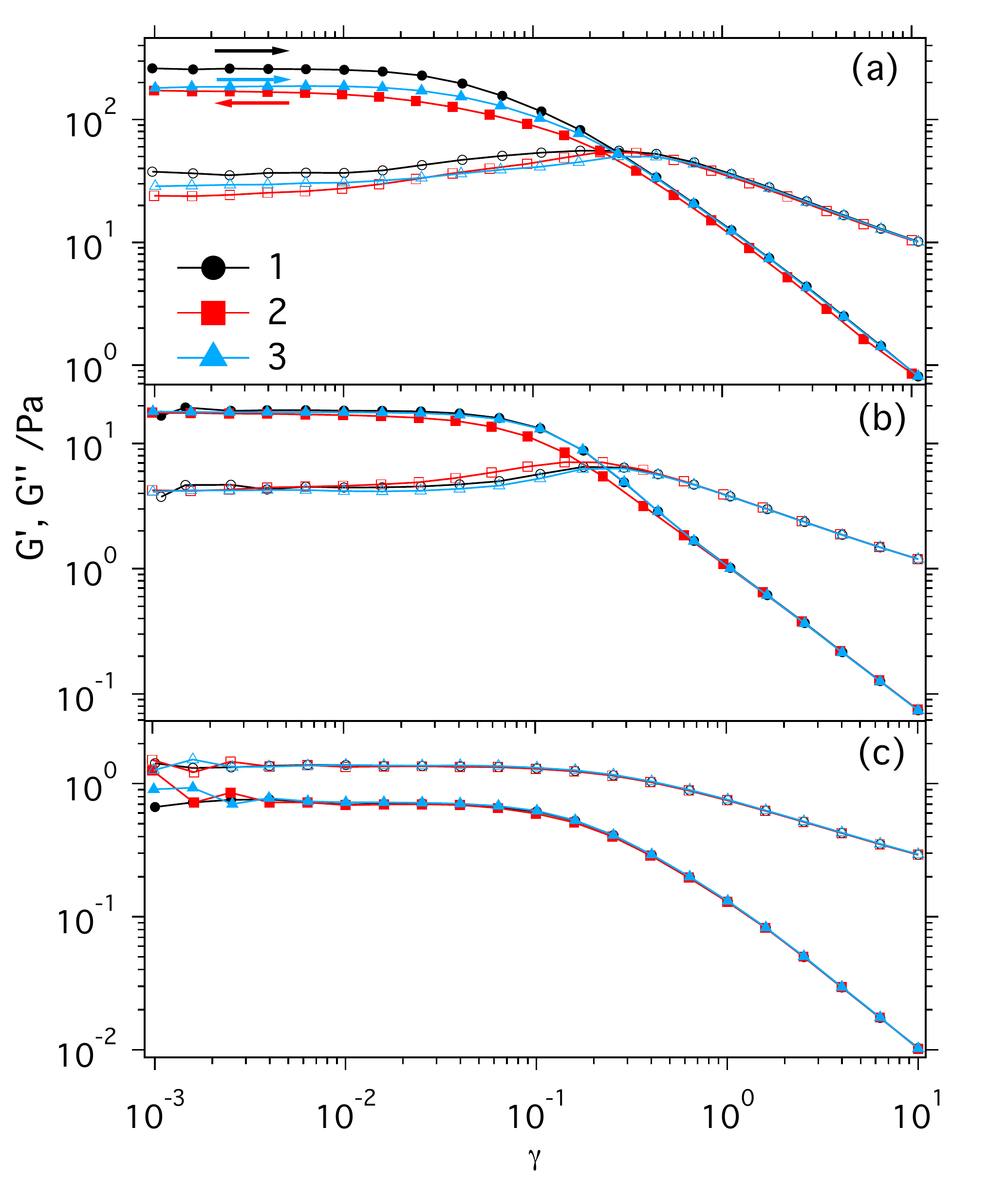}
 \caption{Storage (G$^{\prime}$, closed symbols) and Loss (G$^{\prime\prime}$, open symbols) moduli at $\omega = 1$~rad/s, for the initial Dynamic Strain Sweep (DSS) with increasing strain amplitude $\gamma$ (DSS 1, $\bullet$), a second DSS with decreasing $\gamma$ (DSS 2, $\blacksquare$), and a third DSS with increasing $\gamma$ (DSS 3, $\blacktriangle$), for volume fractions $\phi$: (a) 0.62, (b) 0.58, (c) 0.53. No rejuvenation was performed in between these measurements.}
 \label{fig2}
\end{figure}

\begin{figure}[t]
 \centering
 \includegraphics[scale = 0.6]{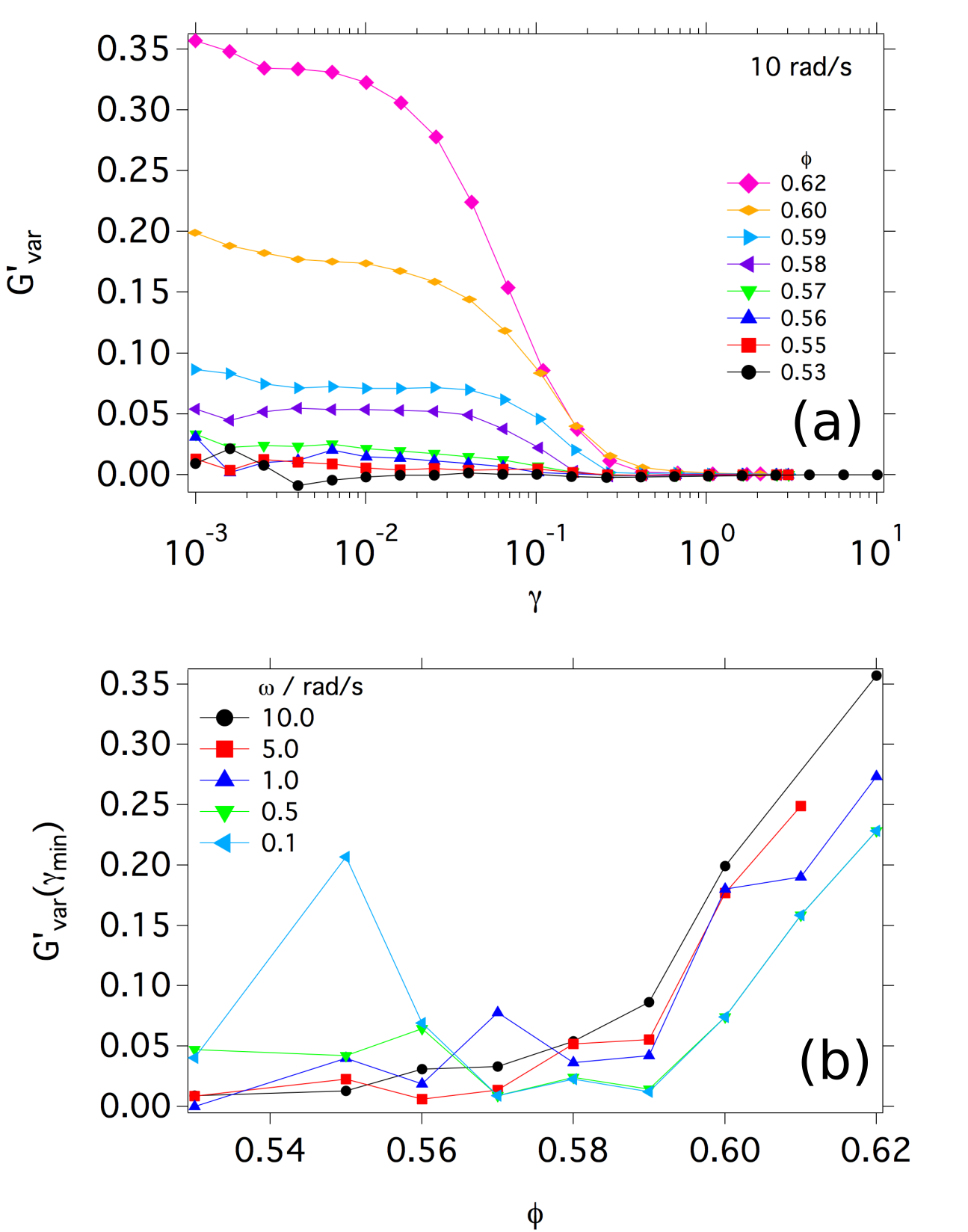}
 \caption{(a) Relative variation of the storage modulus (G$^{\prime}_\mathrm{var}$) with respect to the first DSS, as a function of strain $\gamma$,  for  $\omega = 10$~rad/s, and different volume fractions $\phi$, as indicated. (b) G$^{\prime}_\mathrm{var}$ calculated for the smallest $\gamma$ in (a), as a function of $\phi$, for different frequencies $\omega$, as indicated.}
 \label{fig3}
\end{figure}

For the same samples and oscillation frequencies discussed in the previous section, we compare the first DSS test (DSS 1) with the two successive DSS tests (DSS 2 and DSS 3, Fig.\ref{fig2}). DSS 2 is performed immediately after the first, decreasing $\gamma$ starting from the maximum value reached in the first test. In DSS3 the strain amplitude is increased again using the same protocol as in DSS1. For the fluid at $\phi = 0.53$, the response in DSS 2 reproduces that of DSS 1 (Fig.\ref{fig2}c). This is reasonable for a fluid state in which no history effects are expected. For the sample with $\phi = 0.58$ we observe instead that, while the viscoelastic moduli at large strain amplitudes are again comparable with the first test, at small and particularly intermediate strain amplitudes the moduli are smaller in the second test. This effect becomes much more pronounced for the highest $\phi = 0.62$, where in the linear regime a reduction of approximately 35\% of G$^{\prime}$ is observed. This finding suggests that, in agreement with recent results on similar systems, in which structural changes where monitored under shear \cite{schall_zaccone,ganapathy}, the first fluidisation of the sample induces irreversible structural rearrangements in the glass, which lead to a reduction of the elastic modulus. This is also consistent with the presence of residual stresses after the initial yielding occurs \cite{ballauff,cloitre}.

\begin{figure}[t]
 \centering
 \includegraphics[scale = 0.5]{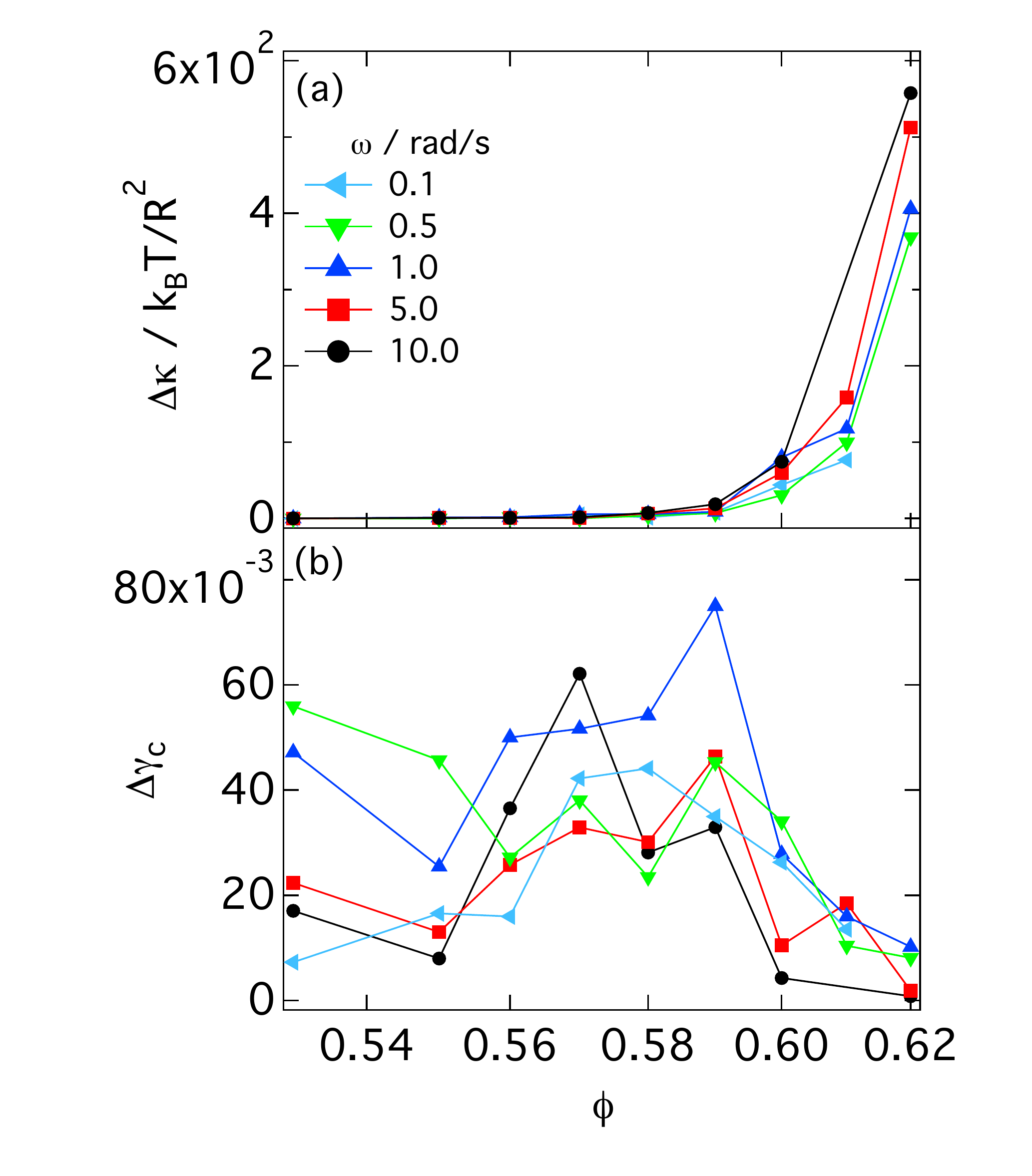}
 \caption{(a) Difference $\Delta\kappa$  and (b) difference $\Delta\gamma_c$, between the $\kappa$ and $\gamma_c$ parameters, respectively, obtained in the fittings of DSS1 and DSS2, as a function of volume fraction $\phi$ and for different oscillation frequencies (as indicated).}
 \label{fig_fitparams_diff}
\end{figure}
%\begin{figure*}[t]
% \centering
% \includegraphics[scale = 0.35]{Fig_lissajous.png}
% \caption{Lissajous figures obtained for sample with $\phi = 0.60$ and  $\omega = 10$~rad/s. Left: Comparison of figures obtained for DSS 1 and DSS 2, for $\gamma$ going from the linear regime to the flow regime, as indicated. Right: Comparison of figures obtained for DSS 2 and DSS 3 , for $\gamma$ in the regime where significant hysteresis is observed in the values of  G$^{\prime}(\gamma)$, as indicated.}
% \label{fig_lissajous}
%\end{figure*}

\begin{figure}[t]
 \centering
 \includegraphics[scale = 0.4]{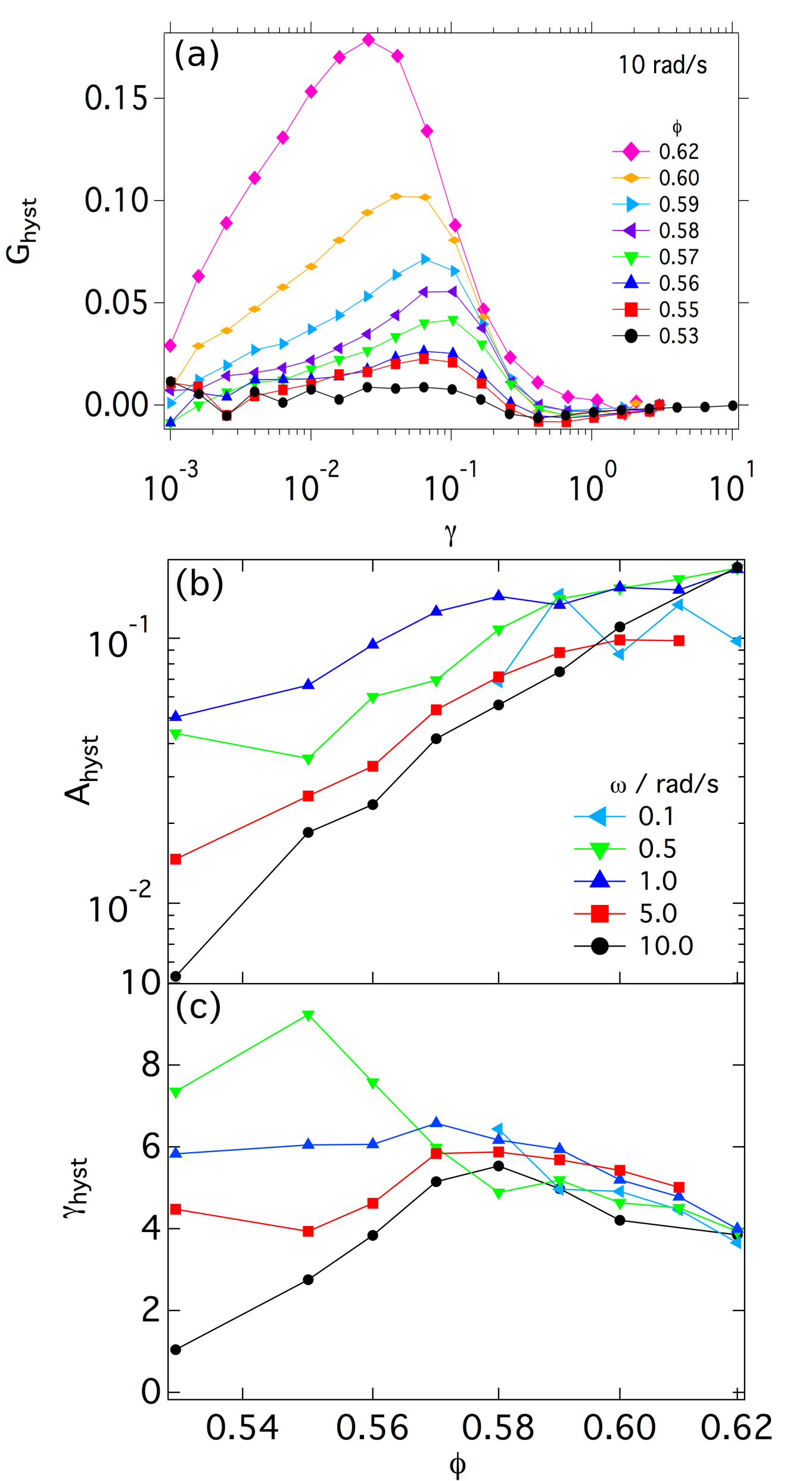}
 \caption{(a) Relative difference in the shear modulus (G$^{\prime}_\mathrm{hyst}$) measured with increasing or decreasing $\gamma$ in the third and second DSS, respectively, as a function of strain $\gamma$,  for  $\omega = 10$~rad/s, and different volume fractions $\phi$, as indicated. (b) Area $A_\mathrm{hyst}$ of G$^{\prime}_\mathrm{hyst}$,  and (c) the value $\gamma_\mathrm{hyst}$ of $\gamma$ related to the maximum hysteresis, as a function of $\phi$, for different frequencies $\omega$, as indicated.}
 \label{fig4}
\end{figure}

The progressive growth of the difference between DSS 1 and 2 with increasing $\phi$ is better visualized through the quantity G$^{\prime}_\mathrm{var}$($\gamma$)=[G$^{\prime}_1$($\gamma$)-G$^{\prime}_2$($\gamma$)]/G$^{\prime\,\,\mathrm{lin}}_1$, where G$^{\prime}_1$($\gamma$) and G$^{\prime}_2$($\gamma$) are the strain-dependent storage moduli measured in the first and second DSS, respectively, and G$^{\prime\,\,\mathrm{lin}}_1$ is the value of the storage modulus in the linear response regime from DSS 1. As shown in Fig.\ref{fig3}a, where G$^{\prime}_\mathrm{var}$($\gamma$) is reported for all measured samples and $\omega = 10$~rad/s, this difference becomes significant and grows fast for $\phi > 0.57$, i.e. entering the glass state. It is also interesting to observe that the difference starts to be significant and grows for $\gamma\lesssim 0.2$, a value of the strain amplitude that has been associated to the cage size in hard-sphere colloidal glasses \cite{pham08}. This supports our interpretation of the difference as the result of structural rearrangements occurring at the level of the cage of nearest neighbours. 
The growth of the difference when entering the glass state is observed at all measured frequencies  and increases with increasing frequency (Fig.\ref{fig3}b).\\
A similar, but less noisy behavior is observed for the difference between the values of the elastic constants obtained by fitting DSS1 and DSS2 with the phenomenological model of Eq.\ref{Gprime} (Fig.\ref{fig_fitparams_diff}a). This confirms that the reformation of a different state in re-solidification after shear-melting is a phenomenon related to the glass state and that becomes increasingly pronounced approaching random close packing. From the model fits of DSS1 and DSS2 we additionally evaluated the difference of the values of $\gamma_c$ obtained in tests 1 and 2 (Fig.\ref{fig_fitparams_diff}b). It apparently shows, within the noise, that a maximum positive difference is observed in the vicinity of the glass transition, i.e. the yielding is observed at significantly larger strain amplitudes in the shear-melting process than in the structure reformation. The microscopic origin of this behavior deserves future investigation.

%We report in Fig.\ref{fig_lissajous} elastic stress vs strain Lissajous figures\cite{mckinley} obtained for DSS 1 and DSS 2 for $\gamma$ values which go from the linear to the flow regime and $\omega = 10$ rad/s. For each DSS the figures show the progressive departure from the elliptic shape characteristic of the linear response, which indicates the onset of non-linearities. At the largest strains the shapes become more rectangular, indicating the transition from an initial linear response, indicated by an approximately linear increase of the stress with strain, to a flow response, corresponding to the region of almost constant stress vs strain. By comparing the results of DSS 1 and DSS 2 one can observe that while the figure in the linear regime at small $\gamma$ and in the flow regime at large $\gamma$ appear comparable, in the intermediate regime of strains non-linear contribution increase more rapidly in DSS2, as indicated by the larger deviation from elliptical shape (see for example data for $\gamma = 0.1$). This is in agreement with what discussed for the harmonic moduli  G$^{\prime}$ and G$^{\prime\prime}$.

We now consider the third DSS, which is measured immediately after DSS 2 and for increasing $\gamma$ (Fig.\ref{fig2}). Again for the fluid sample with $\phi = 0.53$ the results reproduce those of the previous tests. For the sample with $\phi = 0.58$, in the linear regime the response is comparable to DSS 2, but starts to deviate at larger $\gamma$, in the regime where the system yields, approaching the response of DSS 1. At large strain amplitudes $\gamma\gtrsim 1$, all responses overlap. This hysteresis between the third and the second DSS tests becomes even more pronounced for $\phi = 0.62$. The increase of hysteresis with increasing $\phi$ is more precisely quantified through G$^{\prime}_\mathrm{hyst}(\gamma)$=[G$^{\prime}_3(\gamma)$-G$^{\prime}_2(\gamma)$]/G$^{\prime\mathrm{lin}}_3$, where  G$^{\prime}_3(\gamma)$ and G$^{\prime}_2(\gamma)$ are the strain-dependent storage moduli measured in DSS 3 and DSS 2, respectively, and G$^{\prime\mathrm{lin}}_3$ is the average value of  G$^{\prime}_3$ in the linear regime. As shown in Fig.\ref{fig4}a,  G$^{\prime}_\mathrm{hyst}(\gamma)$ shows significant hysteresis when approaching $\phi = 0.58$, similar to G$^{\prime}_\mathrm{var}$($\gamma$), and becomes particularly large for the highest values of $\phi$, deep in the glass state.\\
This trend is confirmed at all frequencies, as shown in Fig.\ref{fig4}b, where the area under G$^{\prime}_\mathrm{hyst}(\gamma)$, $A_\mathrm{hyst}=\int_{-3}^{-1}G^{\prime}_\mathrm{hyst}(10^x)dx$ with $x=\log\gamma$, is plotted as a function of $\phi$ for different frequencies. It is interesting to note that the frequency dependence of the hysteresis is apparently non-monotonic, with maximum hysteresis at the intermediate frequency $\omega = 1$~rad/s. We finally report in Fig.\ref{fig4}c the strain $\gamma_\mathrm{hyst}=10^{\overline{x}}$ obtained from the first moment of G$^{\prime}_\mathrm{hyst}(\gamma)$, $\overline{x} = \int_{-3}^{-1}xG^{\prime}_\mathrm{hyst}(10^x)dx/A_\mathrm{hyst}$. The value $\gamma_\mathrm{hyst}$ is used to estimate the strain at which maximum hysteresis is observed, minimising the noise that would be present if using directly the location of the maximum of data shown in Fig.\ref{fig4}a. In the glass state $\gamma_\mathrm{hyst}$ clearly decreases for all frequencies. On the other hand in the fluid it decreases with $\phi$ at small frequencies but increases at large frequencies. The decrease of   $\gamma_\mathrm{hyst}$ might be associated with the earlier onset of yielding at large volume fractions, as also evidenced for $\gamma_c$ in Fig.\ref{fig6b}. The observation of yielding at smaller strain amplitudes might be associated with the approach to random-close packing and the reduction of the available free volume for deformation.\\

%The presence of hysteresis can be discerned also in the elastic Lissajous figures shown in Fig.\ref{fig_lissajous} (right plot) for DSS 2 and 3 and $\omega = 10$ rad/s. The figures are restricted to the regime of $\gamma$ where hysteresis between forward and backward DSS is observed. The figures of  DSS 2 indicate a more pronounced departure from linear response and a higher contribution of non-linearities at comparable strains, which is coherent with the earlier departure from the linear response regime observed for the storage modulus in DSS 2.

We can qualitatively understand the hysteresis observed at high $\phi$, in the glass state, as follows: In a DSS with increasing strain amplitude (like DSS 1 and 3 in a series),  the sample is initially in a solid state and the maximum amplitude of deformation is progressively increased: the cage opposes elastically the deformation until the strain amplitude is sufficiently large to induce yielding and cage breaking. For even larger strain amplitudes the system starts to flow, with the storage modulus rapidly decreasing with increasing $\gamma$. On the other hand, in a DSS with decreasing strain amplitude (like DSS 2 in a series), the system is initially in a fluid state and by decreasing $\gamma$ starts to reform a solid state with G$^{\prime}$ increasing with decreasing $\gamma$. However, the formation of the solid is partially contrasted at intermediate $\gamma$ by the still undergoing structural disruption induced by the oscillatory shearing. Thus the increase of G$^{\prime}$ with decreasing $\gamma$ for the reforming solid in DSS 2 does not follow the result of a DSS with increasing strain amplitude (DSS 3), it is rather lower due to still ongoing structural disruption. This occurs until the linear response regime is achieved, where the small deformation does not lead to significant structural disruption and the same G$^{\prime}$ is measured for DSS 2 and 3.\\
The emergence of hysteresis in the glass state (and not in the liquid), can also be interpreted from an energy landscape perspective. In the liquid, all minima are shallow and of the same (uniformly low) depth. Hence, over a shearing cycle the system moves from a shallow minimum into another shallow minimum of approximately the same depth, with no appreciable hysteresis. In the glass, instead, the system is initially in a deep glassy minimum (meta-basin) which, in the energy landscape, is far apart from other deep minima of comparable depth. In this case, shear melting brings the system out of the meta-basin into a nearby minimum which cannot be as deep as the original one. This is reflected in the average connectivity being lower in the new steady-state (shallower minimum).
Another manifestation of this scenario, is the widely different anelasticity for liquid and glass, as shown by our data. Anelasticity measures the strength of the departure from the linear response regime to the yielding and flow. For a solid glass originally in a deep minimum, the anelasticity has to be strong (and connectivity change has to be large) to go from the stable initial system into the flowing state. For the liquid, conversely, since the system is initially in a shallow minimum, this change is much smaller. 
 
\subsection{A reproducible shear-induced glass state}

 \begin{figure}[t]
 \centering
 \includegraphics[scale = 0.4]{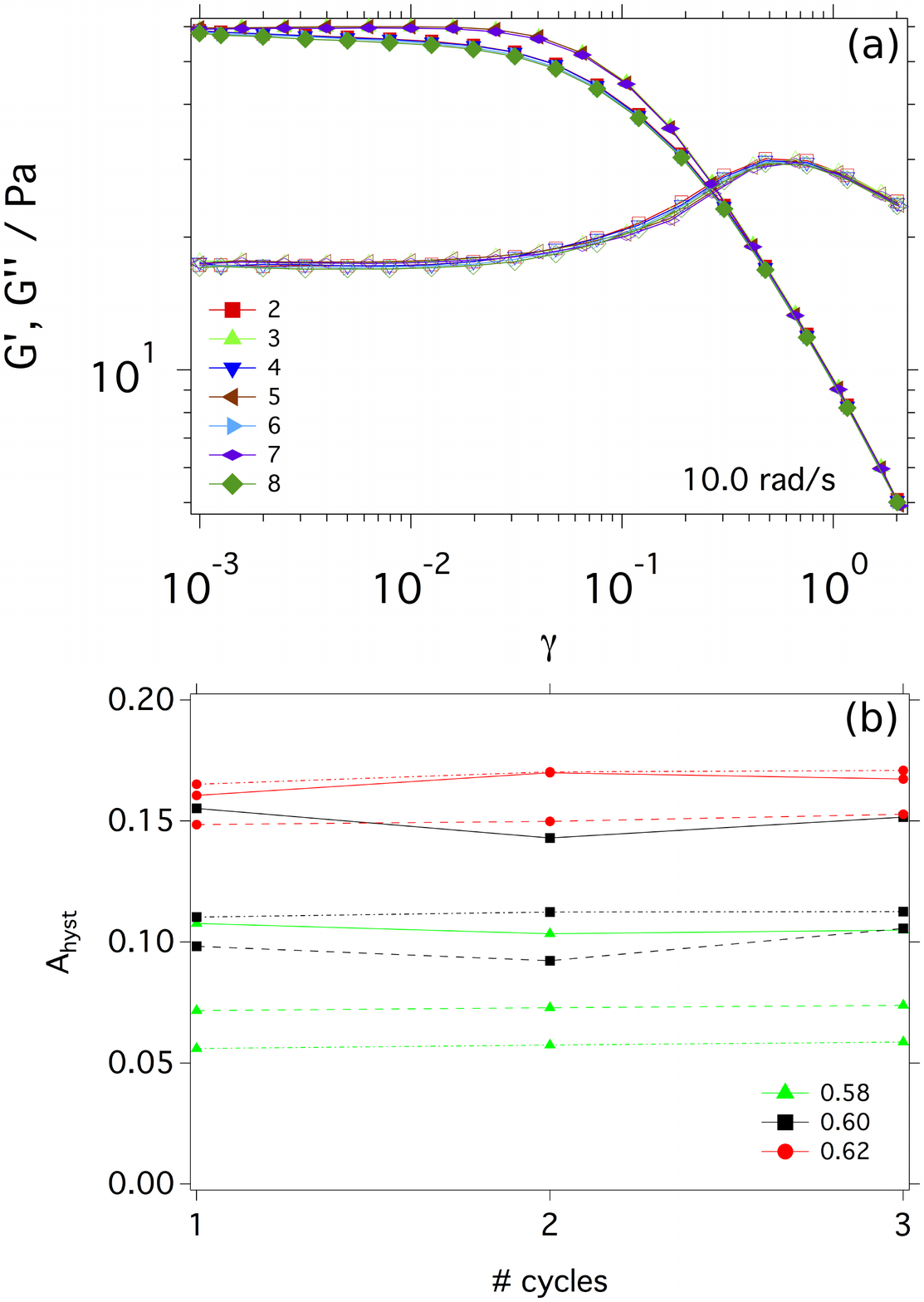}
 \caption{(a) Storage (G$^{\prime}$, closed symbols) and Loss (G$^{\prime\prime}$, open symbols) moduli as a function of strain amplitude $\gamma$, for sample with $\phi = 0.60$ and $\omega = 10$~rad/s, obtained for successive repeated DSS measurements, with the strain amplitude $\gamma$ increasing (odd numbers in legend) or decreasing (even numbers in legend). No rejuvenation was performed in between these measurements. (b) $A_\mathrm{hyst}$ as a function of cycle number, for glassy samples with $\phi = 0.58$, 0.60 and 0.62, as indicated, and frequencies $\omega$ = 0.5 rad/s (solid lines), 5 rad/s (dashed lines) and 10 rad/s (dashed-dotted lines).}
 \label{fig5}
\end{figure}

We extended the repeated measurements of Dynamic Strain Sweeps beyond the third test, alternating tests with increasing and decreasing strain. For all samples, including all glass states, and all frequencies, we find that the response of tests performed with the same direction of strain variation are reproducible, and, for the glass states, the hysteresis between the tests with opposite direction of strain variation remains constant. Data for $\phi = 0.60$ and $\omega = 10$~rad/s are shown in Fig.\ref{fig5}a as an example. This is shown more quantitatively in Fig.\ref{fig5}b, where the area of G$^{\prime}_\mathrm{hyst}$, $A_\mathrm{hyst}$, is presented as a function of melting and re-solidification cycle for 3 samples in the glass state ($\phi = 0.58$, 0.60, 0.62), at different frequencies. No significant variations of $A_\mathrm{hyst}$ are observed for the 3 samples and for the different frequencies in the different cycles (cycle 1: DSS2 and DSS3, cycle 2: DSS4 and DSS5, cycle 3: DSS6 and DSS7). This indicates that, after a transient regime corresponding to the first melting in DSS1 and the successive re-solidification in DSS2, in which residual stresses lead to reformation of a different glass, as shown in the analysis of G$^{\prime}_\mathrm{var}$($\gamma$) in Fig.\ref{fig3}, successive processes of shear-melting and re-solidification become reproducible and the same glass state is obtained repeatedly after re-solidification.\\
 This result is surprising in view of recent studies which show, under application of a constant shear rate or a constant stress, that the flowing dispersion retains residual stresses, which are dependent on the previous shear history and lead to reformation of a different glass state after removal of the shear field \cite{ballauff,cloitre}. It suggests therefore that after the initial yielding, which indeed leads to the re-formation of a different glass, as shown by the different moduli obtained in DSS 1 an DSS 2 both in the linear and non-linear regime, the application of successive melting processes to the post-yielding glass apparently is not resulting in the storage of additional residual stresses. Since these residual stresses were associated to remaining structural deformation in the molten glass \cite{koumakis2016,cloitre}, we can speculate that the consecutive melting processes do not induce any additional persistent structural deformation. This process of training the flowing system by shear is reminiscent of memory effects induced by oscillatory shearing in non-Brownian suspensions, which have been related to a shear-induced structural organization of the particles under shear \cite{keim,foffi,regev2013}. The existence of such shear-induced structural organization in these systems will be the subject of future work.\\
 It is also interesting to note that the reproducibility of the mechanical properties of the glass formed after the initial melting provides a simple and rapid way of rejuvenating a glass \cite{viasnoff}.

%\subsection{Model Results}

\section{Conclusions}

We analysed the process of shear-melting and re-solidification of hard-sphere colloidal glasses, in comparison to concentrated fluids. In particular we investigated the evolution of the first harmonic viscoelastic moduli G$^{\prime}$ and G$^{\prime\prime}$ over a series of consecutive strain sweeps with increasing and decreasing amplitude, at several fixed frequencies. For fluids, we do not observe significant changes in the response along the series. Upon entering the glass instead different effects can be observed. The glass in its initial state after pre-shearing, previous to any oscillatory deformation leading to yielding (pre-yielding glass) can be distinguished from the glass which reforms after a first yielding process (post-yielding glass). The latter presents a smaller storage modulus G$^{\prime}$ in the linear regime, which we can associate to irreversible structural rearrangements of the cage during yielding, in agreement with recent results on silica dispersions \cite{schall_zaccone} and previous results on continuous shearing \cite{ballauff,cloitre}. The post-yielding solid shows a reproducible response over several consecutive strain sweeps, with a reproducible linear modulus, but also the presence of hysteresis effects: the moduli measured with decreasing strain amplitude are smaller  at intermediate strain amplitudes than the corresponding values measured with increasing strain amplitude. We interpret this difference as the result of two distinct physical processes, in which in the strain sweep with increasing $\gamma$ the cage is progressively deformed until it breaks, while in the strain sweep with decreasing $\gamma$ the cage is rebuilt, but under the disturbance of the oscillatory deformation which reduces its resistance. The reproducible glass state obtained after the inital yielding and re-solidification suggests that, after the initial yielding,  successive shear-melting protocols do not lead to the storage of additional residual stresses in the material. These reproducible cycles of shear-melting and re-solidification are reminiscent of memory effects observed in athermal systems \cite{keim,foffi,regev2013}, which lead to a steady structural arrangement of particles over several cycles of deformation. These results indicate also an easy and rapid way of rejuvenating a colloidal glass.\\
We describe the initial melting of the pre-yielding and post-yielding glass in terms of a recently proposed model based on the loss of long-lived nearest neighbors \cite{schall_zaccone,Alessio_new}. The elastic constant extracted from the model shows a strong frequency dependence in the fluid, and weak in the glass. We associate this result with the fact that in the fluid there is a competition between the timescales imposed by shear and associated with Brownian relaxation, while in the glass shear dominates due to the divergence of the timescale of Brownian motion. The model allows to distinguish 3 regimes of deformation for the cage during yielding: a regime of cage flexibility in the fluid, in which the departure from linear response associated with cage rearrangements occurs at increasingly smaller strain amplitudes when approaching the glass transition. This is interpreted in terms of the loss of cage flexibility associated with the slowdown of the dynamics. A second regime around the glass transition in which the cage size is constant and the departure from linear response is independent of $\phi$. Finally a third regime of cage compaction, due to the approach to random close packing, in which again the departure from linear response decreases with $\phi$, due to the decreasing free volume available for deformation. The regime of cage flexibility disappears at high frequencies, since at short timescales compared to the cage relaxation the fluid appears as a solid glass.
% If you have acknowledgments, this puts in the proper section head.
\begin{acknowledgments}
M.L. acknowledges A. B. Schofield for providing the PMMA particles, and the "Direcci{\'o}n de Apoyo a la Investigaci{\'o}n y al Posgrado" of the University of Guanajuato for funding through the "Convocatoria Institucional para fortalecer la excelencia acad{\'e}mica 2015", project title "Statistical Thermodynamics of Matter Out of Equilibrium", as well as funding from "Secretaria de Educaci{\'o}n P{\'u}blica", program "PROMEP", through grant Nr. SEP-23-005.
\end{acknowledgments}

% Create the reference section using BibTeX:
%

\end{document}